\documentstyle[draft,epsf]{mn}

\def\gsim{\mathrel{\raise0.35ex\hbox{$\scriptstyle >$}\kern-0.6em 
\lower0.40ex\hbox{{$\scriptstyle \sim$}}}}
\def\lsim{\mathrel{\raise0.35ex\hbox{$\scriptstyle <$}\kern-0.6em 
\lower0.40ex\hbox{{$\scriptstyle \sim$}}}}
\def\oii{{\rm O{\sc ii}}}

\title{The $K_s$-band Luminosity and Stellar Mass Functions of Galaxies
in $z\sim1$ Clusters}

\author[Kodama \& Bower]
{Tadayuki Kodama$^{1,2,3}$ \& Richard Bower$^3$ \\
$^1$ National Astronomical Observatory of Japan, Mitaka,
Tokyo 181--8588, Japan\\
$^2$ Department of Astronomy, University of Tokyo, Hongo,
Bunkyo-ku, Tokyo 113--0033, Japan\\
$^3$ Department of Physics, University of Durham, South Road, 
Durham DH1 3LE, UK}

\begin{document}

\maketitle

\date{Received, Accepted}

\begin{abstract}
We present the near-infrared ($K_s$-band) luminosity function
of galaxies in two $z\sim1$ cluster candidates, 3C336 and Q1335+28.
A third cluster, 3C289, was observed but found to be contaminated
by a foreground system.
Our wide field imaging data reach to $K_s$=20.5 (5$\sigma$),
corresponding to $\sim$$M^*$+2.7 with respect to the passive evolution.
The near-infrared luminosity traces the stellar mass 
of a galaxy due to its small sensitivity to the recent star formation
history. Thus the luminosity function can be transformed
to the stellar mass function of galaxies using the $J-K$ colours
with only a small correction (factor$\lsim$2) for the effects of
on-going star formation. The derived stellar mass function
spans a wide range in mass from $\sim$3$\times$10$^{11}$~M$_{\odot}$ down to 
$\sim$6$\times$10$^{9}$~M$_{\odot}$ (set by the magnitude limit). The
form of the mass function is very
similar to lower redshift counterparts such as that from
2MASS/LCRS clusters (Balogh et al. 2001) and the $z=0.31$ clusters
(Barger et al. 1998). This indicates little evolution of galaxy
masses from $z=1$ to the present-day.  
Combined with colour data that suggest star formation is completed 
early ($z\gg1$) in the cluster core, it seems that
the galaxy formation processes (both star formation and mass assembly)
are strongly accerelated in dense environments and has been largely completed
by $z=1$. We investigate whether the epoch of mass assembly of massive cluster 
galaxies is earlier than that predicted by the hierarchical galaxy formation 
models. These models predict the increase of characteristic mass by more than
factor $\sim 3$  between $z=1$ and the present day.
This seems incompatible with our data.
\end{abstract}


\section{Introduction}

In the hierarchical structure formation scenario based on a CDM
Universe, a galaxy forms by the assembly of many proto-galactic
fragments. These small systems collapse first in the high density peaks in the
initial density fluctuation fields, and grow by successive mergers
between galaxies.
This overall picture is supported by various pieces of
observational evidence, such as the existence of dynamically decoupled
subgalactic components within a galaxy (eg., de Zeeuw et al. 2002)
or on-going mergers with tidal remains.
Le F\`evre et al. (2000) count the statistical associations
of close companions around galaxies in the redshift survey fields
(CFRS and LDSS) based on the high spatial resolution images by
WFPC2 on the {\it Hubble Space Telescope} (HST). They find a rapid increase
in the fraction of close pairs with redshift, suggesting that a typical
$L^*$ galaxy today has undergone 1-2 significant merger events since
$z\sim1$.
Dickinson et al. (2002) have shown that the most luminous Lyman break
galaxies at $z\sim3$ in the {\it Hubble Deep Field North} are an order of
magnitude less massive than $M_{*}$ today.

Clusters of galaxies are specially selected
areas that grow from high density regions in the large scale structure
of the early Universe. In these regions, the galaxy formation process
is likely to be accerelated. This qualitatively explains why the majority of 
star formation in galaxies in cluster cores take place much earlier
($z\gg1$) than that in the general field (as shown by the extensive
analyses based on the fundamental planes (FP) and the colour-magnitude
relations (CMR) both locally and at higher redshifts
eg., Bower et al. 1992; 1998; Ellis et al. 1997; van Dokkum et al. 1998;
Stanford et al. 1998; Kodama et al. 1998, Kelson et al.2000). 
Compared to star formation, hierarchical galaxy formation 
models (eg., Kauffmann 1996) predict that the mass assembly process is 
more widely spread over time.
Indeed, van Dokkum et al.\ (1999) find many mergers or close companions
in a $z=0.83$ cluster, suggesting that mass assembly is still
in an active phase in this high redshift cluster, even though the star
formation in most of these mergers have been completed (as suggested
from their red colours). The key issue now is
to identify the epoch by which the cluster galaxies have accumulated most
of their mass. This is still an open question.

However, measuring total galaxy masses at high redshifts 
is difficult, since dynamical measurements of internal velocity dispersion
for a large statistical sample becomes extremely time consuming
even with 8-10~m telescopes (van Dokkum \& Stanford 2002).
In contrast, stellar mass can be more easily estimated based on the
near-infrared luminosity.
The near-infrared luminosity of a galaxy is relatively insensitive
to its on-going/recent star formation history even at high redshifts
($z\sim1$), and hence gives a good measure of the underlying total
stellar mass of the galaxy. 
Kauffmann \& Charlot (1998) showed that the evolution of
near-infrared luminosity function (LF) was a sensitive test of galaxy 
formation models, where we should expect to see
strong evolution with redshift if the mass of the galaxies are growing
due to the assembly processes. Kauffmann \& Charlot argued that 
this evolution will be most prominent in the number density of massive 
galaxies, since it takes longer to form these massive systems.

For the brightest cluster galaxies, Arag\'on-Salamanca, Baugh, \&
Kauffmann (1998) have shown that there is significant mass evolution
between $z=1$ and present-day, typically by a factor 2-3. The
observational result that the central galaxy of a $z=1.27$ cluster is
separated to multiple systems supports this view (Yamada et al. 2002).
This can be easily understood, since the brightst cluster galaxy preferentially
sits in the dynamical centre of a cluster, where the satellite
galaxies tend to sink and merge towards the central galaxy due to the
dynamical friction (Binney \& Tremaine 1987).

Pioneering work using the $K$-band LF to quantify the mass evolution
of the cluster galaxy population has been presented by Barger et al. 
(1996; 1998) and De Propris (1999) for many distant clusters out to $z\sim0.9$.
They both find that the change of characteristic luminosities
of the $K$-band LFs with redshift is entirely consistent with the
passive evolution of galaxies formed at high redshifts
($z_f>2$), suggesting mass evolution is negligable between
$z\sim0.9$ and the present. These results have pushed the mass assembly epoch 
of the massive cluster galaxies to higher redshift ($z\gsim0.9$).
It is worth noting, however, that their analyses are limited to the very 
central part of the clusters (typically clustercentric distance of 
$r_c$$\lsim$0.5~Mpc), and may not be representative of the whole cluster 
population due to dynamical mass segregation.

In this paper, we present the combined $K_s$-band luminosity function
of the two $z\sim1$ clusters drawn from a wide field ($r_c$$\lsim$1~Mpc)
in order to study the stellar mass evolution
of the representative cluster population as a whole.
We use $J-K_s$ colours to correct for the effect of star formation on the
mass-to-luminosity ($M/L$) ratio in order to derive the stellar mass. 
The crucial issue is whether there is a deficit of massive galaxies
($<M^{\ast}$) at $z\sim1$ compared to the less massive galaxies.
A detection of this deficit would confirm the importance
of galaxy mergers, as suggested by hierarchical galaxy formation models.

An outline of the paper follows. After describing the
cluster selection, observation and the data reduction in \S~2, 
and the data processing including the corrections for incompleteness
and foreground and background contamination in \S~3,
we present the $K_s$-band number counts, luminosity function, and the
stellar mass function of the $z\sim1$ clusters in \S~4.
Discussion and conclusions are given in \S~5 and \S~6, respectively.
The cosmological parameters adopted throughout this paper are
($h_{100}$, $\Omega_0$, $\Lambda_0$)=(0.7, 0.3, 0.7),
unless otherwise stated.
Here we define $h_{100}$ and $h_{70}$ as $H_0$/(100~km s$^{-1}$
Mpc$^{-1}$) and $H_0$/(70~km s$^{-1}$Mpc$^{-1}$), respectively.

\section{Observation and Data Reduction}

\begin{table*}
\caption{Summary of the observations}
\label{tab:obs}
\begin{center}
\begin{tabular}{ccccccc}
\hline\hline
Cluster  & Redshift & RA (J2000) & Dec (J2000) & $J$ (sec)  & $K_s$ (sec) & FoV (arcmin$^2$) \\
\hline
3C336    & 0.927    & 16 24 39.1 & 23 45 13    & 5,400      & 10,320 & 12.56\\  
3C289    & 0.967    & 13 45 26.4 & 49 46 33    & 5,400      &  5,520 & 12.94\\  
Q1335+28 & 1.086    & 13 38 07.5 & 28 05 11    & 5,400      & 10,440 & 13.09\\  
\hline
\end{tabular}
\end{center}
\end{table*}

\subsection{$z\sim1$ cluster sample}

We selected three cluster candidates at $z\sim1$, 3C336, 3C289 and
Q1335+28, which are all associated with QSO's. The positions and
redshifts of the QSO's are given in Table~1. Each of our sample is a
highly plausible cluster at similar redshift, $z\sim1$.  The existence
of the 3C336 cluster was firmly established by Steidel et al.\ (1997)
in their spectroscopic observations of the galaxies around the QSO to
identify the QSO absorbers.  Bower \& Smail (1997) detected a weak
lensing signal in the HST WFPC2 images towards this field which is
probably originated from the deep cluster potential associated to the
QSO.  The 3C289 cluster also shows a merginal weak lensing signal
(Bower \& Smail 1997).  Best (2000) has shown a clear number excess of
galaxies around the 3C289 QSO in their near-infrared image and the HST
optical image. These galaxies form a tight red colour-magnitude
sequence, indicating the existence of a cluster.  The Q1335+28
(B2~1335+28) cluster was first noted by Huchings et al. (1995) as the
excess of \oii\, emitters in a narrow-band survey.  Yamada et
al. (1997) have confirmed an excess of red galaxies which form a
colour-magnitude sequence.  Tanaka et al. (2000) conducted a detailed
photometric analysis of this cluster based on the optical and
near-infrared imaging data.

\subsection{Observation}

We have conducted a wide-field, near-infrared imaging in $J$ and $K_s$ bands
of the three $z\sim1$ clusters with INGRID on the 4.2m telescope
{\it William Hershel Telescope} (WHT) at La Palma for 28-31 Mar 2001.
Our exposure times are given in Table~1.
INGRID has a large field of view of 4$'$$\times$4$'$ with a pixel scale of
0.238$''$.
The actual field of view of the combined frames are slightly smaller
due to the dithering and are given in Table~1. The typical field size is
3.6$'$$\times$3.6$'$,
covering 1.7$h_{70}^{-1}$~Mpc on a side at the cluster redshifts.
The field centres are chosen at the QSO positions, except for
the Q1335+28 cluster where the field centre is shifted to 1$'$ to
the South and 1$'$ to the West to optimally cover the known structures
of the cluster (Tanaka et al. 2000).

The second half of the run had good seeing conditions
at FWHM$\sim$0.6-0.7$''$. The combined frames have FWHM=0.7-0.8$''$.
The observation was mostly photometric and the photometric zero-points
were calibrated using the UKIRT faint standard stars (Casali \&
Hawarden; Hawarden et al. 2001) taken during the photometric nights.
The 5$\sigma$ limiting magnitudes are $J$=23.0 and $K_s$=20.5, which
correspond to $\sim$$M^*$+2.5-2.7 with respect to passively evolving
galaxies at $z=1$.

\subsection{Data reduction}

We used {\sc IRAF} software in reducing the data following standard procedures.
Flat fielding was made using the super-flat constructed by combining
all the non-aligned frames. We then subtract the sky background by median 
filtering. Finally, we median combine the frames after aligning
the frames using 2-5 stellar objects.
SExtractor v.2.1.6 (Bertin \& Arnouts 1996)
is then used for object detection in the $K_s$-band frame.
More than 5 connected pixels with counts greater than 1.5 $\sigma$
above the background noise level is detected as an object.
$J$-band magnitude is measured at each point of the $K_s$-band
selected object.
The SExtractor output MAGBEST is used as an estimate of the total
magnitude in $K_s$, and the 2$''$ diameter aperture is used to derive the
colour in $J-K_s$.
Some obvious stars are excluded on the basis of the SExtractor output
(CLASS\_STAR$>$0.6) in the $J$-band (which is 0.1-0.3 mag deeper
than $K_s$ for the passive cluster members).
Colour images of the three systems are shown in Figs.~1--3.

\begin{figure*}
\begin{center}
  \leavevmode
\vspace*{4cm}
3c336\_colour.jpg
\vspace{4cm}
\end{center}
\caption{
A false-colour picture of the 3C336 cluster created from
our $J$ and $K_s$ images.
The field size is 3.54$'$$\times$3.55$'$.
North is up and East is to the left.
}
\label{fig:3c336}
\end{figure*}

\begin{figure*}
\begin{center}
  \leavevmode
\vspace*{4cm}
3c289\_colour.jpg
\vspace{4cm}
\end{center}
\caption{
A false-colour picture of the 3C289 cluster created from
our $J$ and $K_s$ images.
The field size is 3.58$'$$\times$3.61$'$.
North is up and East is to the left.
}
\label{fig:3c289}
\end{figure*}

\begin{figure*}
\begin{center}
  \leavevmode
\vspace*{4cm}
q1335\_colour.jpg
\vspace{4cm}
\end{center}
\caption{
A false-colour picture of the Q1335+28 cluster created from
our $J$ and $K_s$ images.
The field size is 3.62$'$$\times$3.61$'$.
North is up and East is to the left.
}
\label{fig:q1335}
\end{figure*}

\section{Data Analysis}

\begin{figure*}
\begin{center}
  \leavevmode
  \epsfxsize 0.48\hsize
  \epsffile{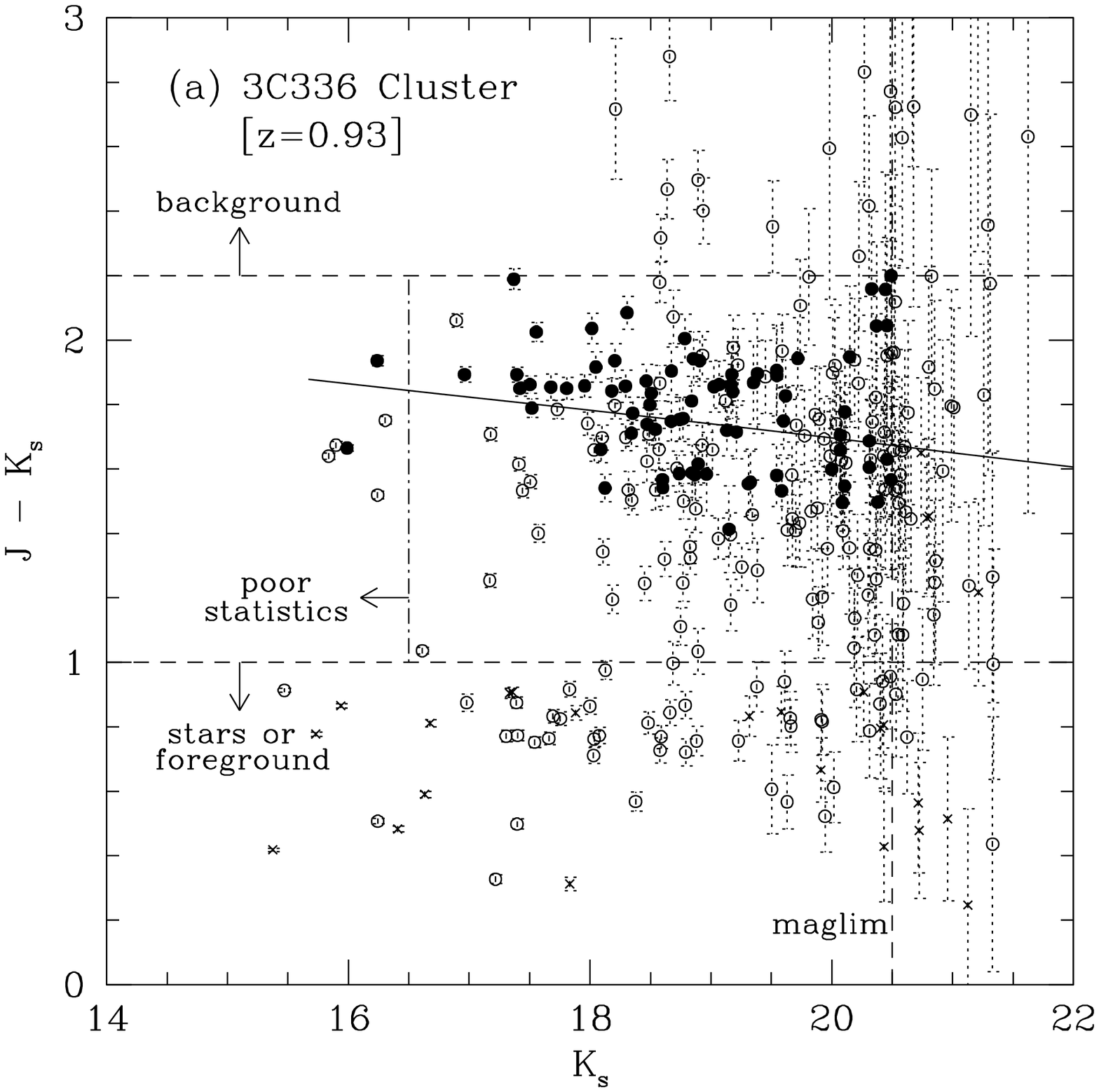}
  \epsfxsize 0.48\hsize
  \epsffile{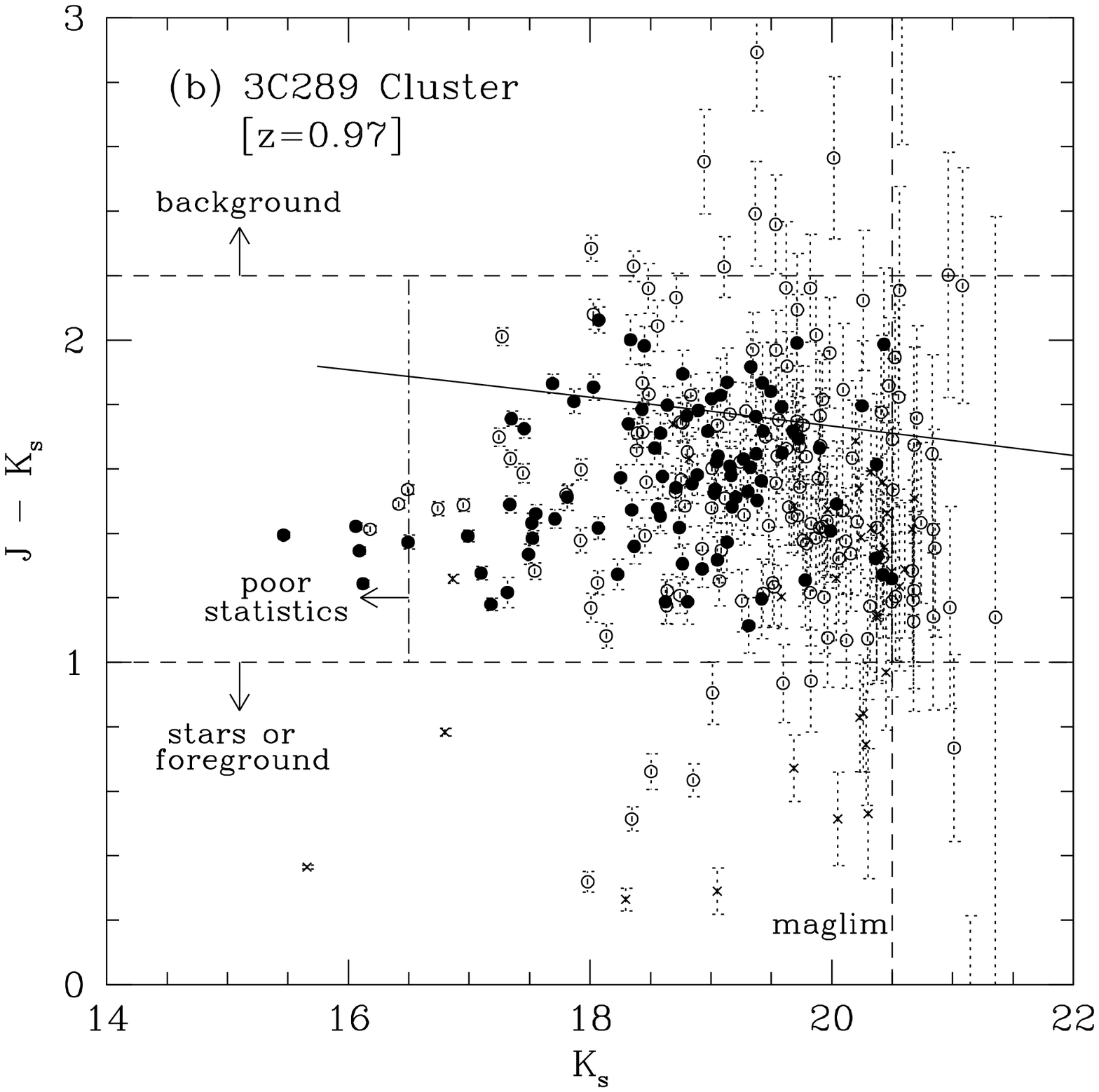}\\
  \hspace*{-0.1cm}
  \epsfxsize 0.48\hsize
  \epsffile{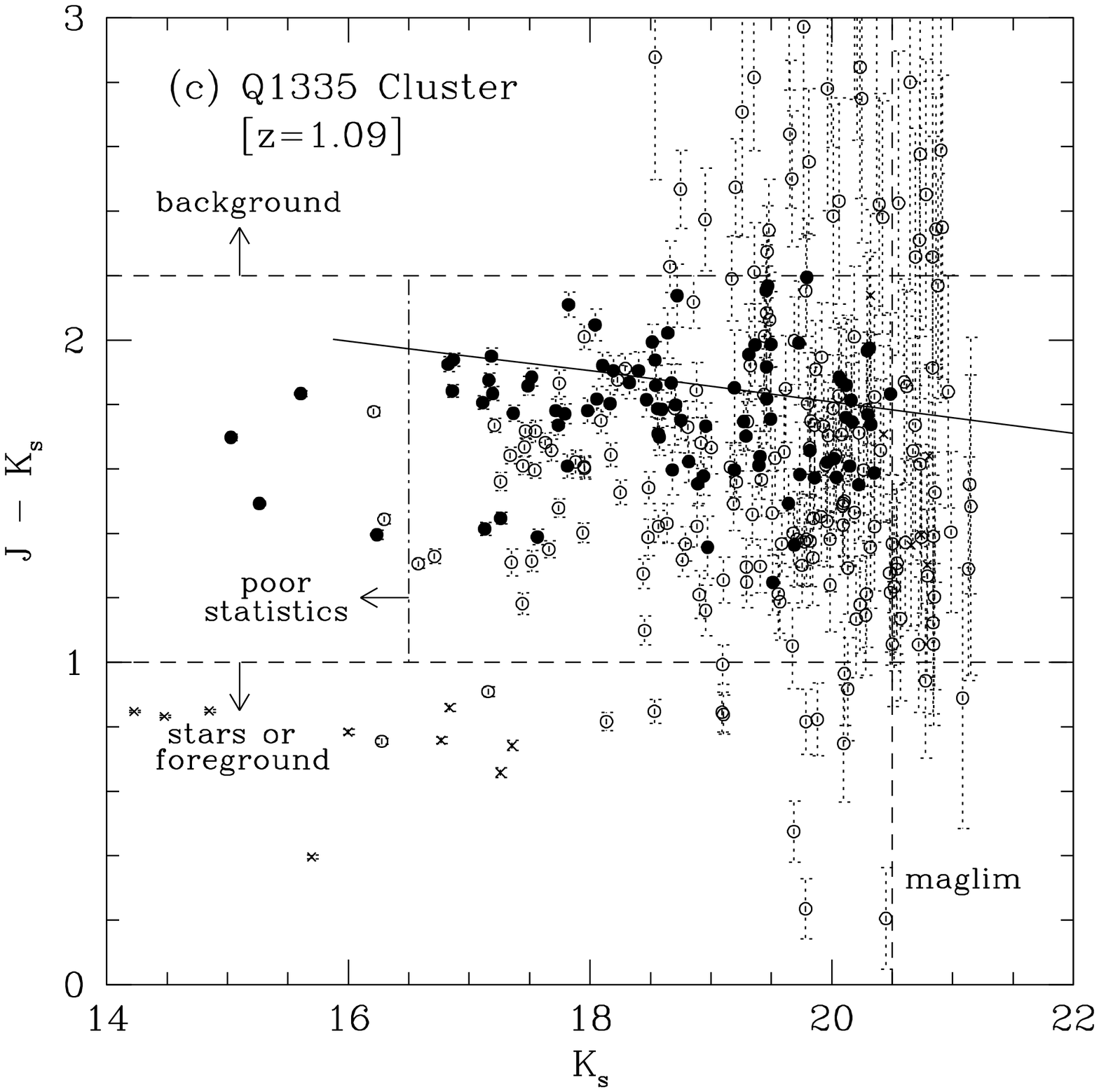}
  \epsfxsize 0.48\hsize
  \epsffile{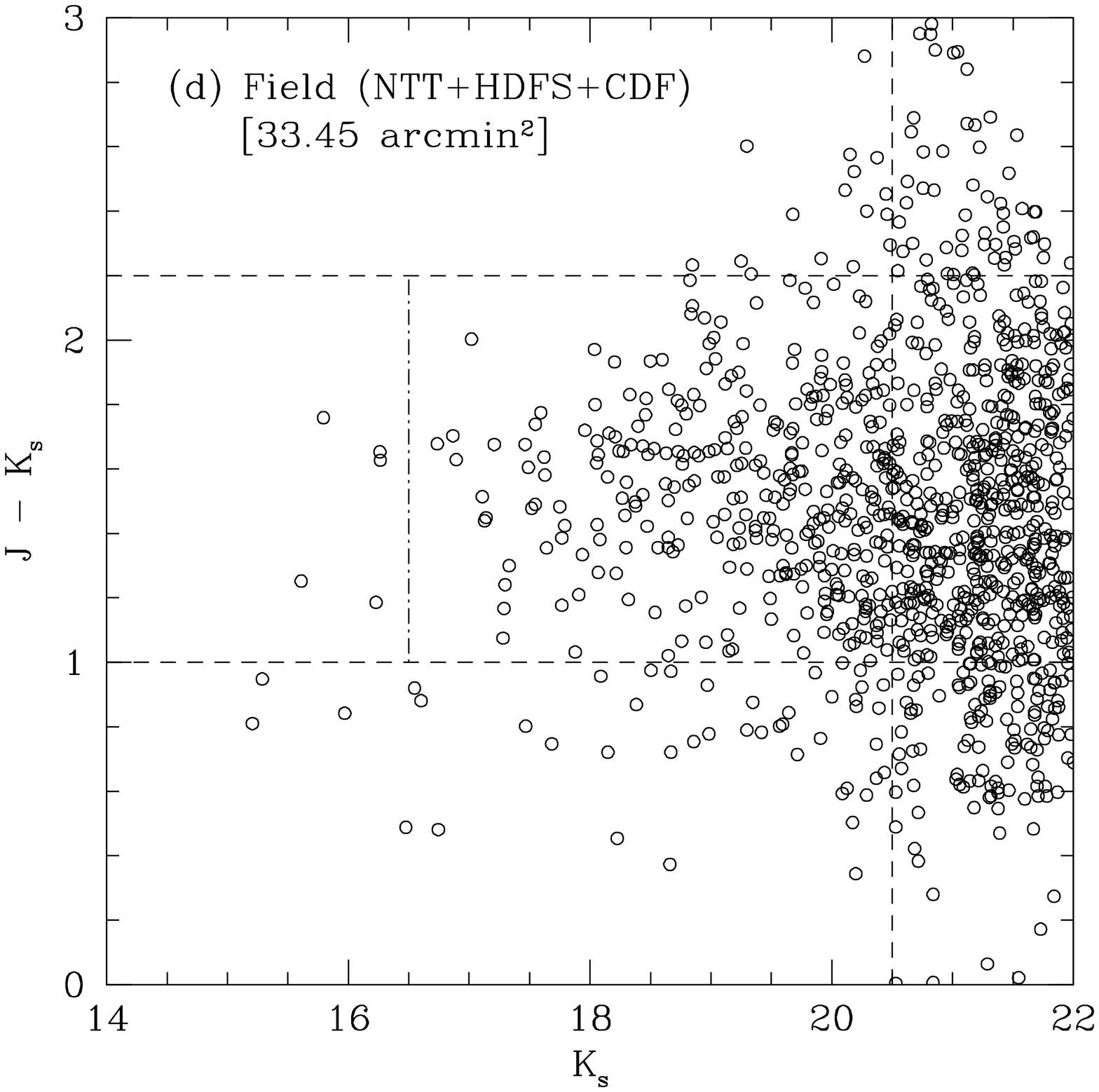}
\end{center}
\caption{
Colour-magnitude diagrams for the three $z\sim1$ clusters and
the combined blank fields (Saracco et al. 1999; 2001).
Note that the field of view of the combined blank fields is
approximately 2.6 times larger than that of each cluster field.
The solid lines represent the loci of the colour-magnitude relations
at cluster redshifts predicted by the passive evolution models ($z_f$=5)
normalised to Coma cluster (Kodama et al. 1998).
The filled circles indicate the `statistical' members from a typical
Monte-Carlo run of the field subtraction.
The crosses are possibly stars, and excluded from the analyses.
The perpendicular dashed line shows the magnitude limit in our analyses
(5~$\sigma$).
The horizontal dashed lines (upper and lower) and the perpendicular
dot-dashed line outline the limits beyond which the galaxies are 
excluded from further analysis due to large background contamination,
contamination by stars and foreground, and poor statistics 
in the field subtraction, respectively (see text for details).
}
\label{fig:cmd}
\end{figure*}

\subsection{Completeness correction}

Since we are interested in the number density of the faint objects,
it is essential to correct for the completeness for the object detection
at faint magnitudes.
To estimate the completeness, we use `artificially dimmed true images',
rather than artificially generated galaxies, which are embedded in
the data frames.
We first construct a `noise frame' by combining the randomly shifted
individual frames instead of being aligned correctly according to the
dithering (as done for constructing the `science frame' in \S~2.3).
We then dim the properly combined original science frame by 1-2 magnitudes.
We now combine the noise frame and the dimmed science frame together,
by properly weighting the two frames according to the dimmed magnitudes.
In this way, we can make up a new frame where the objects are dimmed
by a certain magnitudes while keeping the noise amplitude unchanged.
The object detection using the SExtractor is then processed
on this frame using the same set-up parameters as used for the original
frame, and count the number of galaxies which are detected in the
original science frame but are not detected in the dimmed new frame.
From this simulation, we estimate the completeness to be
$\sim$90\% at the $K_s$=18.5-19.5 bin, $\sim$80\% at $K_s$=18.5-19.5,
and $\sim$70\% at $K_s$=19.5-20.5 magnitudes for 3C336 and Q1335+28 clusters.
For 3C289, the completeness is slightly lower due to the lower exposure
time in the $K_s$-band ($\sim$90\% at $K_s$=16.5-17.5,
$\sim$80\% at $K_s$=17.5-18.5, $\sim$70\% at $K_s$18.5-19.5
and $\sim$50\% at $K_s$=19.5-20.5), but this cluster will be excluded
from our sample to construct the luminosity and mass functions due to
foreground contamination (see \S~3.2).
We scale the galaxy counts by multiplying the inverse of the completeness
factor at each magnitude bin. The maximum factor that is multiplied
for the incompleteness correction is therefore only 1.429 (or 0.155 dex)
at the faintest bin ($K_s=19.5-20.5$) for 3C336 and Q1335+28 clusters.
In the colour-magnitude diagrams in Figs.~\ref{fig:cmd}a--c,
we generated additional faint galaxies to compensate the incompleteness by 
scattering the existing galaxies according to the Gaussian photometric errors.

\subsection{Stars, foreground and background galaxies subtraction}

Our star and field galaxy subtraction scheme has two steps.
Firstly, a loose colour cut in $J-K_s$ is applied to initially suppress
the contamination. And secondly, a statistical field subtraction
is applied using the blank field data.

In addition to the star exclusion based on the CLASS\_STAR (\S~2.3),
we use a colour cut at $J-K_s<1.0$ (Figs.~\ref{fig:cmd}) to suppress
the remaining star contamination.
The objects bluer than this line are likely to be stars or
blue foreground galaxies, since these colours are too blue for
cluster members. For example, a constant star formation model with
$z_f=5$ gives $J-K_s$$\sim$1.1 at $z=1$.
In contrast, most stars are bluer than $J-K_s$=1.0 except for
M giants which can be as red as $J-K_s\sim 1.3$
(Bessell \& Brett 1988).
Therefore, the $J-K_s$ cut is an effective way to subtract many of
the stars and blue foreground populations while keeping the cluster
members.

We also apply a $J-K_s>2.2$ colour
cut (Figs.~\ref{fig:cmd}) well above the red colour-magnitude
sequence of the cluster members at $z\sim1$
($\gsim$2$\sigma$ photometric errors at our magnitude limit)
in order to suppress contamination from red background galaxies
and to thus reduce the statistical fluctuations in the background subtraction.
Although we cannot deny the possibility that some member galaxies could
be redder than this limit due to the heavy dust extinction even in 
$J-K_s$ ($A_V\gsim2$; Calzetti et al. 2000), these colours are much redder
than passively evolving cluster members that track the reddest colour
envelope. 

With the above colour cuts, and the additional magnitude cut at $K_s$=20.5
applied, we then apply a statistical field subtraction based on the blank 
field data.
Here we use the deeper wide-field near-infrared data-set from the NTT Deep
Field, the Chandra Deep Field, and the Hubble Deep Field South
(Saracco et al. 1999; 2001). The total area amounts to 33.45~arcmin$^2$,
significantly larger area than our individual cluster field
($\sim$13~arcmin$^2$).
Figures~\ref{fig:cmd} show a typical Monte-Carlo run of the statistical field
subtraction for each cluster. We divide the colour-magnitude diagram under 
consideration into 6$\times$4 bins (within the colour and magnitude cuts) 
with the steps of 1 magnitude in $K_s$ and 0.25 magnitude in $J-K_s$.
The details of the method should be referred to Appendix~A in
Kodama \& Bower (2001).
The filled circles represent the plausible cluster members
that are statistically in excess above the blank field.

The 3C336 and Q1335+28 cluster candidates show clear sequences of red 
galaxies near the passive evolution prediction at each cluster redshift
(solid lines in Figs.~\ref{fig:cmd}), confirming the existence of the massive
clusters. There are some very bright galaxies ($K_s$$>$16.5) that are
statistically selected as members. However, judging from their
unrealistically bright magnitudes, these are likely to be foreground
galaxies which are retained due to the low number statistics in this
colour-magnitude range. In fact, there are only a few galaxies
brighter than $K_s$=16.5 both in the cluster fields and the blank field,
and it is very hard to statistically remove the field contamination
correctly in this range.

As shown in Fig.~\ref{fig:cmd}b, 3C289 cluster has a large number of
blue galaxies ($1.2\lsim J-K_s\lsim 1.5$) left after the field
subtraction process, as well as many red galaxies on or near
the passive evolution line at the cluster redshift.
Judging from the brightness and the narrowness of the colour distribution
of these blue excess galaxies, these are likely to come from a foreground
structure(s) in excess of the average field rather than blue cluster members.
Therefore, we do not include this 3C289 cluster in the following
analyses in order to avoid the possible impact of this foreground 
contamination.

There are also a few bright blue galaxies with $K_s<18.0$ and $J-K_s<1.6$
in the remaining two clusters, especially in the Q1335+28 cluster,
that are identified as cluster members on the basis of statistical
field subtraction. It would be surprising if these galaxies were
genuine cluster members, given their blue colours for their brightness.
Such bright, blue galaxies are not seen in spectroscopically confirmed
sample of cluster members in the lower-$z$ clusters studied by
such as MORPHS (eg., Dressler et al. 1999) and CNOC (eg., Yee et
al. 1996) projects. Because of this uncertainty,
in the following analyses, we show the results for both cases where
these bright, blue galaxies are included, and excluded.

\section{Results}

\subsection{The $K_s$-band Number Counts}

\begin{figure}
\begin{center}
  \leavevmode
  \epsfxsize 1.0\hsize
  \epsffile{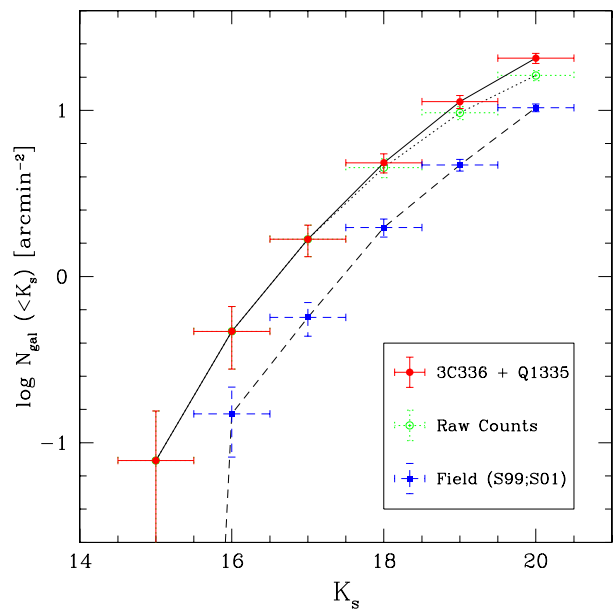}
\end{center}
\caption{
Integrated galaxy number counts.
The objects bluer than $J-K_s=1.0$ are excluded due to a large
contamination of stars.
The dotted and the dashed curves show the raw number counts
in the two cluster fields (3C336 and Q1335+28) and the blank fields,
respectively. The solid curve represents the incompleteness corrected
number counts for the cluster fields.
The error bars are based on the simple Poisson statistics.
We see a factor 2-3 excess in the number density of galaxies
in the cluster fields compared to the blank fields. 
}
\label{fig:count}
\end{figure}

Figure~\ref{fig:count}
shows the integrated number counts of galaxies in $K_s$-band
for our cluster fields (dotted line for the raw counts and solid line
for the incompleteness corrected one) and for the blank field (dashed line).
The error bars represent the simple Poisson errors.
Both counts show smooth increase towards fainter magnitudes,
and the excess of the counts in our cluster field compared to the blank
field is clear and is order of factor 2-3 depending on the magnitude
range that is integrated over. This excess supports the existence of the 
distant clusters in these QSO fields.

\subsection{The $K_s$-band Luminosity Function}

\begin{figure}
\begin{center}
  \leavevmode
  \epsfxsize 1.0\hsize
  \epsffile{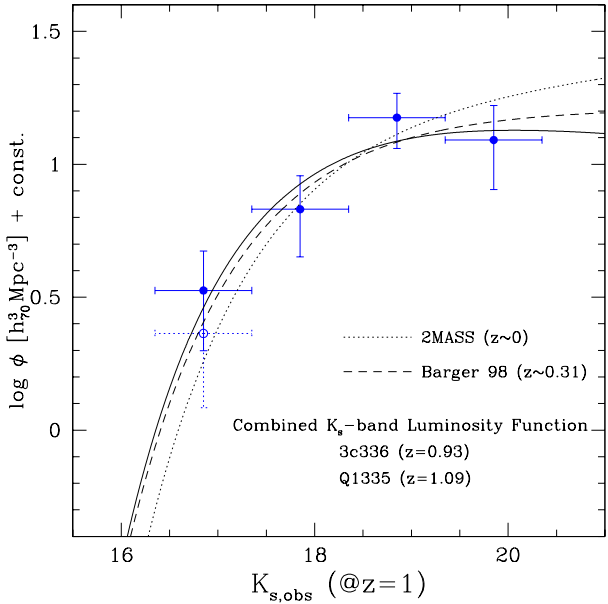}
\end{center}
\caption{
The $K_s$-band luminosity functions in the observed frame seen at $z=1$.
The filled circles with the error bars and the solid curve represent the
combined luminosity function of the two clusters (3c336 and Q1335+28)
and the Schechter function fit to the data with a fixed faint-end
slope of $\alpha=-0.9$.
The open circle with dotted error bars shows the case where the bright blue
galaxies ($K_s<18$ and $J-K_s<1.6$) are excluded. Corrections for
incompleteness and field contamination have been applied.
The error bars indicate the Poisson statistics based on the number of cluster
galaxies and that of the subtracted field galaxies.
The dotted and the dashed curves indicate the luminosity functions
for the lower redshift clusters taken from Balogh et al. (2001) (2MASS/LCRS)
and Barger et al. (1998) ($z\sim0.31$), respectively.
These lower-$z$ functions are transformed to $z=1$ according to the
passive evolution ($z_f=2$), and the normalisation is taken so that
all the curves have the same amplitude at each $K_s^*$.
Note that, {\it after passive evolution has been taken into account,} 
the $K_s$-band luminosity (hence stellar mass) function has hardly
changed since $z\sim1$.
}
\label{fig:klf}
\end{figure}

\begin{figure}
\begin{center}
  \leavevmode
  \epsfxsize 1.0\hsize
  \epsffile{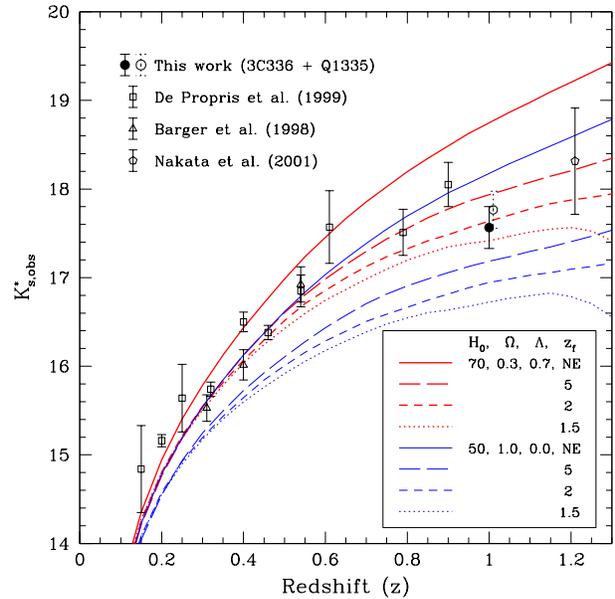}
\end{center}
\caption{
The characteristic magnitudes ($K_s^*$) in the observed $K_s$-band are 
plotted against cluster redshift.
The normalisation is set at $z=0$ using the 2MASS/LCRS luminosity function
for clusters (Balogh et al. 2001).
The adopted faint-end slopes ($\alpha$) in the fits are $-1.0$ for Barger
et al. (1998) and $-0.9$ for all the others.
The filled circle represents the $K_s^*$ of this work at $z\sim1$,
while the open circle shows the one derived if the bright blue galaxies
($K_s<18$ and $J-K_s<1.6$) are excluded.
Both points are consistent with the passive evolution model with
$z_f\sim2$ in the $\Lambda$-Universe.
}
\label{fig:kstar}
\end{figure}

We now construct the $K_s$-band LF of the $z\sim1$ cluster galaxies to
investigate the mass distribution of cluster galaxies at this high
redshift.  We combine the two clusters 3C336 and Q1335+28 to reduce
the statistical uncertainty. We note that the individual LFs of these
two clusters have consistent properties within errors.

The completeness correction at the faint magnitudes is applied as in
\S~3.1. The field contamination is also subtracted using the same blank
field data as in \S3.2, but here we do not use the colour information
more than to restrict the galaxies to $1<J-K_s<2.2$,
and simply subtract the field counts in each magnitude bin, instead of
doing it by a Monte-Carlo simulation in each colour-magnitude grid (\S~3.2),
to make this process simpler.
The $K_s$-band LF of each cluster is transformed to the common redshift
of $z=1$ by correcting both for the small k-corrections and the small distance
differences ($\Delta$$K_s$=+0.12 and $\Delta$$K_s$=$-$0.15 in total for 3C336
and Q1335+28, respectively), and then combined.

The filled circles with error bars in Fig.~\ref{fig:klf}
show the combined LF of the two clusters
in the observed $K_s$ band seen at $z=1$.
The solid line indicates the best-fit Schechter function.
The error bars represent the Poisson statistics based on the number of
cluster galaxies and that of the subtracted field galaxies.
The open circles show the case where the bright and blue galaxies are
excluded from the analysis with the suspicion that these galaxies
are not cluster members.
Since the faint-end slope ($\alpha$) of the Schechter function is not
well determined by the data due to its depth, the slope is fixed to
$\alpha=-0.9$, the same slope used in De Propris et al.\ (1999)
who took this value from their Coma data (De Propris et al.\ 1998).
We compare this $z\sim1$ LF to the lower redshift counterparts
such as 2MASS/LCRS clusters at $z\sim0$ (Balogh et al.\ 2001) and
the $z\sim0.31$ clusters by Berger et al. (1998) by transforming
the low redshift data to $z=1$ according to passive evolution with $z_f=2$. 
This shown by the dotted and the dashed curves respectively. 
The normalisation of is set so that
all the curves have the same amplitude at each $K_s^*$.  As can be seen, 
once passive evolution is taken into account, further evolution of the 
mass function is not detected.

This is also clearly shown in Fig.~\ref{fig:kstar} which indicates
the evolution of the observed $K_s^*$ of the LF as a function of cluster 
redshift, compared
to the passive evolution models with different formation redshifts,
and no-evolution models for two different cosmologies.
The filled circle represents the $K_s^*$ in the Schechter curve in
Fig.~\ref{fig:klf} and the open circle shows the one if the bright blue
galaxies ($K_s<18$ and $J-K_s<1.6$) are excluded.
Both of these points are fully consistent with the passive
evolution models with $z_f>1.5$ in both cosmologies.
Therefore, the magnitude change in the $K_s$-band LF since $z\sim1$ to the
present-day can be fully descibed by pure luminosity evolution
following the passive evolution expected in an old stellar system.

We note however that luminosity evolution and the stellar mass
evolution is degenerate in this figure (eg., van Dokkum \&
Stanford 2002).
For example, if the formation redshift were significantly younger,
eg. $z_f<1.5$, the observed $K_s^*$ at $z\sim1$ clusters would be
fainter than that predicted by the passive evolution and hence some
stellar mass evolution (in the sense of having less massive systems at
 high redshifts)
would be allowed. In this case, however, the galaxy colours would be
significantly bluer than those observed.
The passive evolution of early-type galaxies with $z_f$=1.5 would give 
$J-K_s=1.5$ at $z=1$, which would be already significantly bluer than the
majority of the likely members in $z\sim1$ clusters (see Fig.~\ref{fig:cmd}).
Therefore by combining the luminosity evolution and the colour
evolution, we can simultaneously put constraints on the formation
redshift of the cluster galaxies and their mass evolution.
In this way, we argue that both the star formation and the mass
assembly processes of galaxies have been largely completed by
$z\sim1$ in cluster cores. We can reinforce this arguement by estimating
the stellar masses of individual galaxies from their $K_s$-band magnitudes
and the $J-K_s$ colours and deriving the stellar mass function
directly. This is discussed in the following section.

\subsection{The Stellar Mass Function}

\begin{figure*}
\begin{center}
  \leavevmode
  \epsfxsize 0.48\hsize
  \epsffile{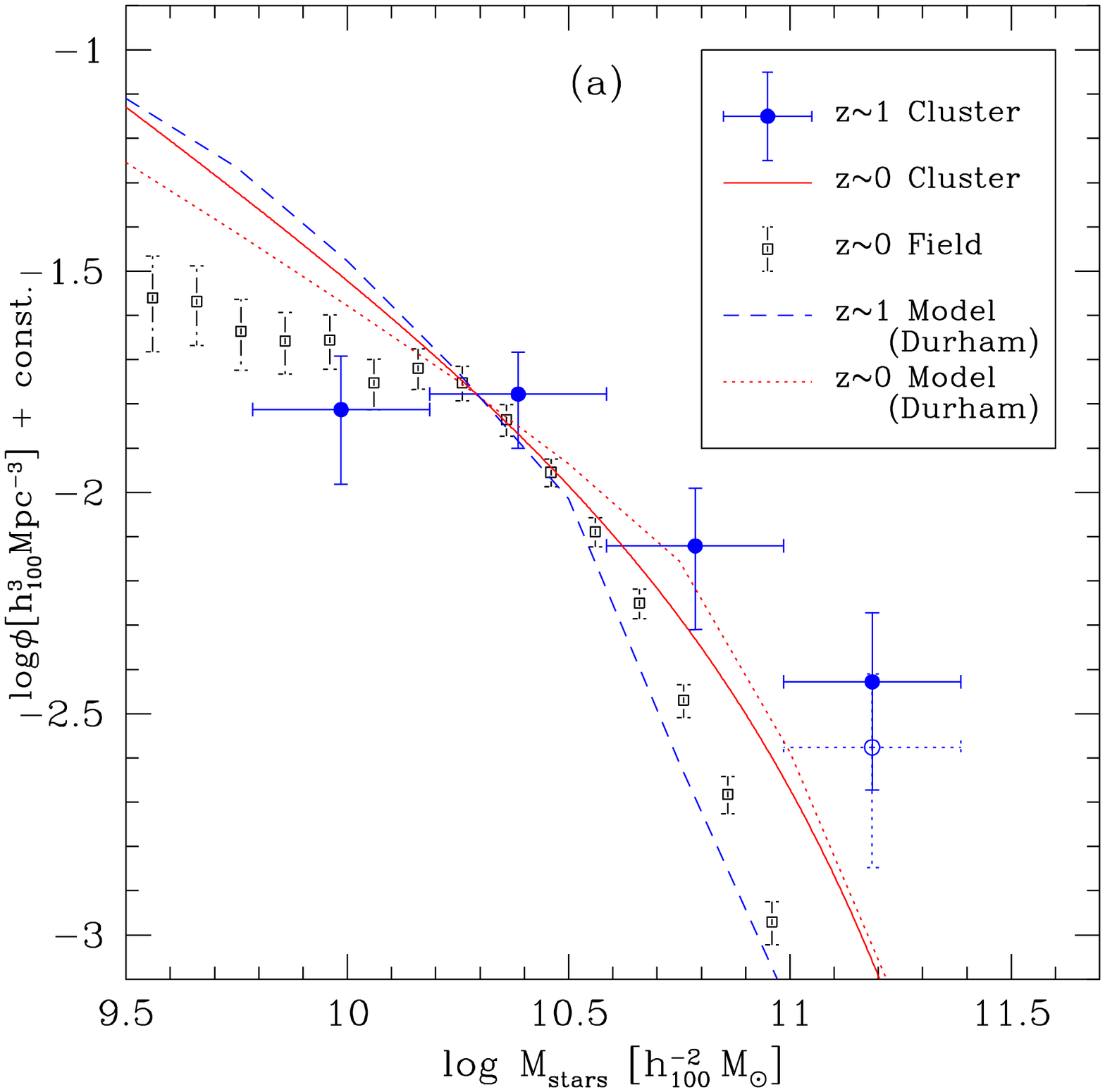}
  \hspace{0.5cm}
  \epsfxsize 0.48\hsize
  \epsffile{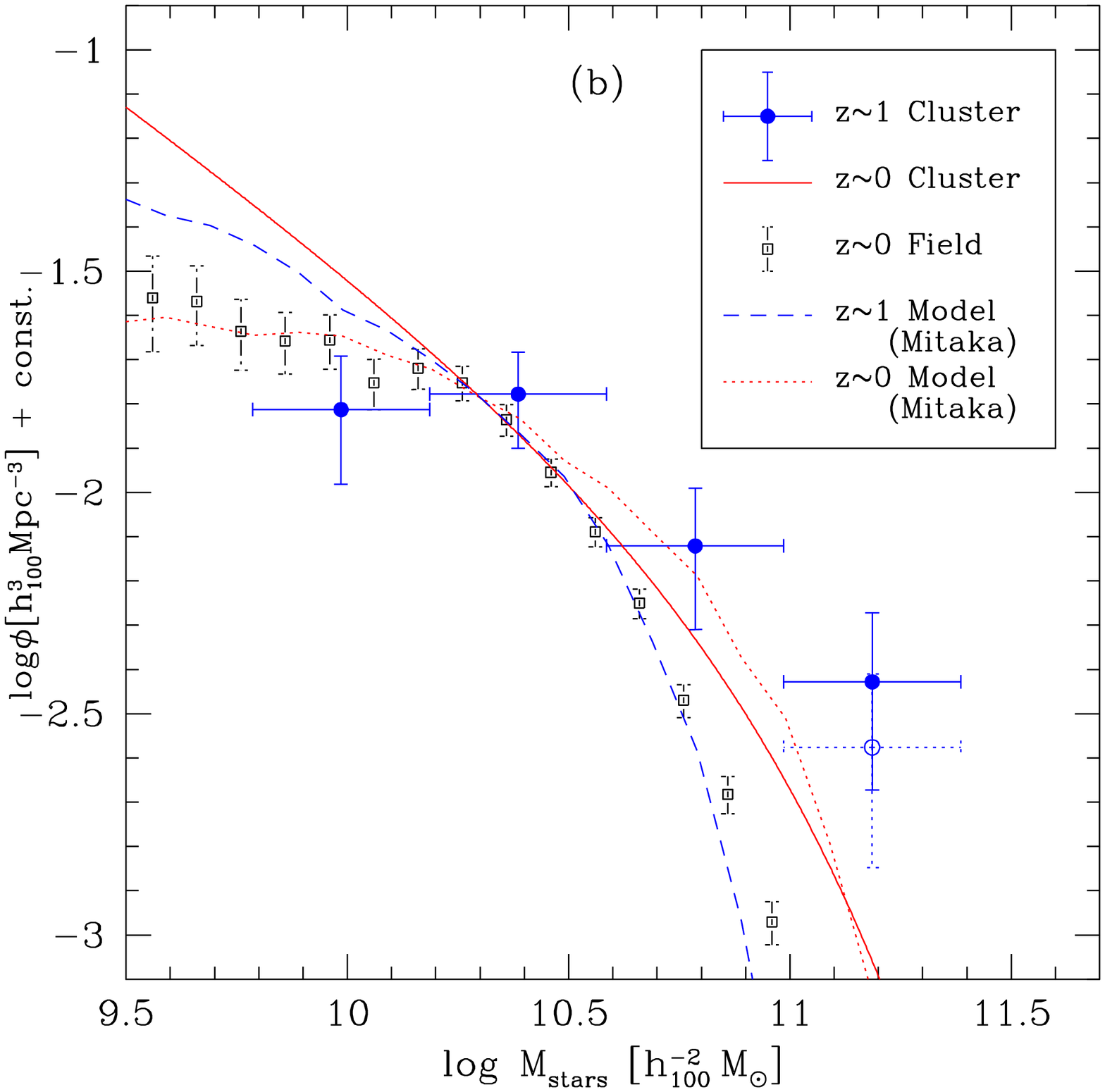}
\end{center}
\caption{
(a) The stellar mass functions.
The filled circles with the errorbars represent the
combined stellar mass function of the two clusters (3C336 and Q1335+28)
corrected for the star formation effect using the $J-K_s$ colours of
individual galaxies.
The open circle with dotted errorbars shows the case where
the bright blue galaxies ($K_s<18$ and $J-K_s<1.6$) are excluded.
The errorbars indicate the Poisson statistics based on the numbers of cluster
galaxies and those of the subtracted field galaxies.
The squares with errorbars and the solid curves show the local stellar
mass function of 2dF galaxies (Cole et al.\ 2000) and that of galaxies
in the 2MASS/LCRS clusters (Balogh et al.\ 2001).
The dotted and the dashed curves represent the Durham semi-analytic model's
predictions of galaxy stellar mass functions for galaxies in $z=0$
clusters and for those in $z=1$ clusters, respectively
(Cole et al.\ 2000).
In this comparison, we adopt the Kennicutt's (1983) IMF and the
($H_0$, $\Omega_0$, $\Lambda_0$)=(70, 0.7, 0.3) cosmology for
consistency among all the data.
The mass functions based on the observed data and the models
are all normalised at 2$\times$10$^{10}$M$_{\odot}$ so as to
have the same amplitude as the 2dF one. Even with this 
conservative normalisation, the observed mass function for the $z\sim1$ 
clusters shows an excess of massive galaxies compared to the model
prediction.\, \,
(b) The same as (a), but the Mitaka models are plotted instead of the Durham 
ones.
}
\label{fig:mf}
\end{figure*}

The $K_s$-band luminosity approximates the stellar mass of a galaxy
due to its relatively small sensitivity to recent star formation
activity and dust extinction. However, in order to derive an accurate
stellar mass of an individual galaxy, we should apply a small
correction to the $M/L$ ratio depending on the galaxy's star formation
history. This correction can decrease the derived stellar mass by upto
factor$\sim$2 for vigourously star forming galaxies.

The recent star formation activity of each galaxy with respect to
its underlying passively evolving component affects not only on the
stellar $M/L$ ratio in the $K_s$-band but also the $J-K_s$ colour
because both reflect the stellar population mix of the galaxy.
Specifically, the larger the recent star formation activity is,
the smaller the $M/L$ ratio is and the bluer is the $J-K_s$ colour.
Therefore by establishing the relation between $M/L$ ratio and
$J-K_s$ colour, we can more reliably transform the $K_s$ magnitude
into the stellar mass by using the known $J-K_s$ colour of an individual
galaxy.
(We note that the `stellar mass' refered to in this paper takes into 
account the mass loss from evolved stars and novae.
The evolution of $M/L$ ratio in the population synthesis model that is
used to transform luminosity to stellar mass includes this effect,
as does the stellar mass predicted by the semi-analytic models
used later in \S~4.3.)

We take the relation between these two quantities from the models in a
sequence of bulge-to-total ratios (Kodama, Bell, Bower 1999),
where the disk component (which is mimicked by the exponentially declining
star formation with a time scale of $\tau$=3.5~Gyr and an infall time scale
of $\tau_{\rm in}$=3.5~Gyr) is added onto the underlying
passively evolving bulge component (which is represented by the elliptical
galaxy model that has an old initial star-burst within the first Gyr).
Using a different set of models (such as taking a different age for the
underlying bulge or adding a secondary burst to the bulge)
do not significantly change the relation between the $M/L$ ratio and the
$J-K_s$ colour since both quantities always change in the same direction.

In this analysis, we shift the zero-point in the observed $J-K_s$
colours by +0.07 magnitude for the 3C336 cluster, since the observed red
sequence is bluer than the model sequence (Fig.~\ref{fig:cmd}a).
This zero-point mismatch can be probably atributed to a model
uncertainty and/or an observational error on the absolute photometry.
We note, however, that this mismatch does not significantly
affect the conclusions presented in this paper, since it corresponds to
the $M/L$ difference of only $\sim$10$\%$.

Furthermore, when we compare our mass function with the other data from 
the literature, we apply a correction to our mass function for 
differences in the IMF and cosmology. We have derived our mass functions 
with power-law IMF with $x=1.1$ and with the Hubble constant of $H_0$=70, 
whereas the other data presented use Kennicutt's (1983) IMF and $H_0$=100.
These differences shift the derived mass by 0.2 and 0.31 dex, respectively.
The mass in our analysis have been therefore decreased by these amounts
in Figs.~\ref{fig:mf} and Table 2 to compare with the other data. 

The combined stellar mass function of the
galaxies in the two $z\sim1$ clusters is presented in Figs.~\ref{fig:mf}.
The open circle shows the case where the blue and bright galaxies are
excluded (as in Fig.~\ref{fig:klf}).  The solid line indicates the 2MASS/LCRS
result (Balogh et al.\ 2001) who derived the stellar mass function of
galaxies in local clusters from the $K$-band LF by applying the star
formation correction using the D$_{4000}$ index that is equivalent to
a colour.  The 2dF local stellar mass function is also shown by the
squares. In the absence of an independent constraint on the total
mass of each cluster, the normalisation of the mass function is arbitrary.
The curves have therefore been renormalised to 
the same amplitude at 2$\times$10$^{10}$M$_{\odot}$. Over the range spanned 
by our data, the shapes of our mass function at
$z\sim1$ and the 2MASS/LCRS or 2dF ones at $z\sim0$ look very similar.
This indicates again that the mass evolution is little since $z\sim1$
to the present-day. We quantify this comparison below by comparing the 
observed mass function with those predicted from semi-analytic models.

\subsubsection{Comparison with the semi-analytic models}

The fact that the massive galaxies ($>$10$^{11}$M$_{\odot}$) are already
abundant in the $z\sim1$ clusters can potentially put a very strong
constraint on the hierarchical galaxy formation models.
In such a models, massive galaxies are the end products of the
repeated mergers of the smaller galaxies, and hence take a relatively
long time to form. We compare our results with semi-analytic models
from Durham and Mitaka (Cole et al.\ 2000 and Nagashima et al.\ 2002,
respectively)
prepared for us by Carlton Baugh and Masahiro Nagashima respectively.
The Durham model prediction is made for
the haloes with the mass chosen so that the space density of the haloes
is the same as that of the present-day rich clusters such as Abell clusters
(2$\times$10$^{-6}$$h^{-1}$ M$_{\odot}$ Mpc$^{-3}$), whereas
the Mitaka model (Nagashima et al.\ 2002) prediction is made for the
galaxies in the 100 haloes whose
circular velosity $V_c$ is equal to $\sim$1000~km/s chosen at $z=1$,
The Durham and Mitaka models are independent implementations of the
semi-analytic scheme, and the details of the treatment differ. In particular,
the Mitaka models allow for mergers between satellite galaxies while the 
Durham models do not (see Cole et al.\ and Nagashima et al.\ for 
more discussion). The models differ significantly in the faint end slope 
of the luinosity function that is predicted, but the overall comparison with
the observational data leads to similar conclusions, as we show below.

Both models predict rapid evolution in the position of the knee of the
luminsoity function, even though the clusters have been selected in a 
way that mimics the observational data. However, the normalisation of
the mass function is free because the absolute numbers of galaxies in
each system depends on the total halo mass. Unfortunately we are not able
to accurately constrain the total halo mass and thus cannot derive an
independent normalisation. We have therefore normalised the stellar mass 
functions so that all the curves have the same amplitude at 
2$\times$10$^{10}$M$_{\odot}$.
As a result of this normalisation we concentrate only on the shapes
of the stellar mass functions. As can be seen from Figs.~\ref{fig:mf}
(dashed line), 
the Cole et al.\ (2000) and the Nagashima et al.\ (2002) models
seem to predict a much steeper mass function 
than is observed, and thus under-predicts the numbers of massive
galaxies in distant clusters. 

\begin{table}
\caption{The giant-to-dwarf ratio for the observed data (2MASS/LCRS and
this work with/without the bright blue galaxies)
and the Durham and Mitaka semianalytic models (Cole et al.\ 2000;
Nagashima et al. 2002).
The error on the ratio for the observed value ($z\sim1$ clusters)
corrresponds to the Poisson statistics, and the one for the Mitaka
semianalytic model represents the scatter among the 100 generated clusters.}
\begin{tabular}{ll}
\hline\hline
cluster ($z$)    & $R$(giant/dwarf)\\
\hline
2MASS/LCRS clusters ($z\sim0$) & 0.13 \\
This work ($z\sim1$) & 0.35$\pm$0.12 \\
This work ($z\sim1$) with bluecut & 0.32$\pm$0.11 \\
Durham model ($z=0$) & 0.18 \\
Durham model ($z=1$) & 0.07 \\
Mitaka model ($z=0$) & 0.21 \\
Mitaka model ($z=1$) & 0.07$\pm$0.05 \\
\hline
\end{tabular}
\label{tab:ratio}
\end{table}

To quanify the comparison between the models and data in a way that is 
independent of the normalisation, we define the giant-to-dwarf 
ratio, $R$(giant/dwarf), as
\begin{equation}
R{\rm (giant/dwarf)}=\frac{n_{\rm gal}(10.6<\log M_{\rm stars}<11.4)}
{n_{\rm gal}(9.8<\log M_{\rm stars}<10.6)}
\end{equation}
The ratios are calculated from data shown in Figs.~\ref{fig:mf},
and are summarised in Table~2.
This shows that the fraction of giant galaxies at $z=1$ in the
semi-analytic predictions is factor $\gsim$3 smaller than the observed
value.
The scatter of the ratio in the Tokyo model measured from the 100 generated
clusters (0.05) is too small to account for the over-abundance of the
massive galaxies in the real high redshift clusters.

How much mass evolution can be allowed between $z=1$ and the present-day
within the errors of our analysis?
To answer this question, we shift the local mass function of 2MASS/LCRS
clusters (Balogh et al. 2001) towards lower masses and calculate the
statistical deviation from the observed mass function at $z\sim1$
by a boot-strap resampling method (ie., randomly replacing the observed points
assuming the Gaussian errors on each point).
From this analysis, we find that we can reject a mass decrease of a 
factor of 2 at 99\% level (2.5$\sigma$) no matter whether the bright blue
galaxies are included or excluded.
Since the definition of bright blue cut that we introduced in \S4.2
is somewhat arbitrary, we consider the effect of further reducing the 
numbers of galaxies in the brightest bin, moving it 
two sigma down from its original place. Even in this extreme case, the 
giant-to-dwarf ratio is 0.27
and is still 2.3 $\sigma$ off from the factor 2 evolution in mass,
corresponding to 98\% rejection level.

\subsection{Less massive galaxies}

In contrast to the bright cluster galaxies,
faint galaxies seem to have a much greater diversity of formation histories.
The faint galaxies may be in general formed at later cosmic times than 
the massive galaxies, although this appears to be opposite to
the hierarchical picture.
If so, then distant clusters should have a LF
that has a declining faint-end slope in contrast to the rising faint-end
slope of local clusters (Balogh et al. 2001; De Propris et al. 1998).
The progression of formation activity to lower-mass galaxies as the
Universe ages is often referred to as the ``down-sizing'' effect
(Cowie et al.\ 1996; Kodama \& Bower 2001).
Kajisawa et al.\ (2000) suggested just such a deficit of faint galaxies
($K>20$, or $M^{\ast}$+1.5) in the 3C324 cluster at $z=1.21$.

In our two $z\sim1$ clusters, however, such a strong deficit is not confirmed
down to $K_s=20.5$ ($M^*$+2.7) (Fig.~\ref{fig:klf}).
A marginal discrepancy ($\sim$2$\sigma$) between $z=0$ and 1 is seen
in the least massive bin of the stellar mass functions (the solid lines
and our data points in Figs.~\ref{fig:mf}).
However, given the fact that
the faint-end slope of the local mass function of cluster galaxies
(2MASS/LCRS; Balogh et al.\ 2001) is only poorly determined
($\Delta\alpha\sim0.3$ (1$\sigma$)), we cannot discuss further whether
there is an evolution between $z=0$ and 1 in the faint end of the
mass function from the current data set.
To test the hypothesis of different formation epoch depending on galaxy
mass, our data would need to go deeper  than $K_s^*>20.5$.
However, we certainly see significant numbers of less massive galaxies 
below 1.5$\times$10$^{10}$M$_{\odot}$, which are nearly absent in
Kajisawa et al.\ (2000).
We speculate that the reason why they saw such a deficit of less massive
galaxies is their limited field coverage
(cluster-centric distance $r_c$$<$40$''$ or 0.33~Mpc).

One might be concerned that our high giant-to-dwarf ratio at $z=1$
clusters and hence the relatively strong constraint on the mass growth rate,
is partly driven by the deficiency of the less massive galaxies at $z=1$
as well as the abundance of massive galaxies.
We verify this effect on our conclusions by artificially
increasing the numbers of galaxies in the least massive bin at $z=1$
by 2$\sigma$, and by flattening the faint end slope of the $z=0$ curve
(2MASS/LCRS) by $\Delta\alpha=0.3$.
However, these effects on the constraint on the mass growth factor is found 
to be small.
For example, in the first case, the giant-to-dwarf ratio goes down to
$R$=0.26 and 0.24 with and without the bright blue galaxies, respectively,
but the mass
evolution of factor 2 between $z=1$ and 0 (which gives $R$=0.07) is still
rejected at 98\% level (2.3$\sigma$).

\section{Discussion}

It is now generally accepted that the stellar populations of galaxies
in clusters are old with little on-going star formation based on the colours
and the fundamental planes studies (eg., Bower et al., 1992; 1998;
Ellis et al. 1997; van Dokkum et al. 1998; Stanford et al. 1998; Kodama et al.
1998; Kelson et al. 2000).
However, the stellar populations of galaxies may be old, but at the 
same time the galaxies may not have been completely assembled. Between $z=1$ 
and the present-day, galaxies might grow in mass through collisions.
This key issue can only be tested by comparing the luminosity functions
of galaxies in clusters at different redshifts.

Early assembly of massive galaxies is a challenge for hierarchical models
of galaxy formation since these models tend to grow galaxies through
both on-going merging and gas accretion up to the present-day. The
first generation of such models predicted very rapid galaxy growth 
(Kauffmann \& Charlot 1998). More recent models within a $\Lambda$-CDM 
universe  and revised merging prescriptions (eg., Cole et al., 2000)
have predicted evolution that is slower but still significant --- see
Cimatti et al.\ (2002) for a recent discussion and comparision with field
data.

In clusters of galaxies, we might anticipate that the rate of evolution
of galaxy mass is less than in the field since rich clusters are over
dense regions of the universe in which gravitational collapse has been
accelerated. 
Thus an important issue that has concerned us is that clusters which are
observed at $z\sim1$ are not directly equivalent to the progenitors of 
clusters in the local universe
(ie., while the galaxies in the distant clusters form part of the local
cluster population, there are many galaxies in local clusters which were
located in lower density environments at $z=1$). Care is therefore need to 
select galaxies from the semi-analytic models, and we have tried to select 
clusters in the same way in both the semi-analytic models and the data.
We applied two different ways of selecting the systems in the models ---
by mass and by space density --- but both definitions give similar results
and predict a significant reduction in the characteristic stellar mass
of cluster galaxies with redshift.

In contrast to the model, however, our observations of two
distant clusters show that the
evolution of the luminosity function is well modelled by passive 
evolution of the galaxy population. The $K$-band luminosity function 
is slightly brighter at $z=1$ than at $z=0$. However, when we
correct the observed galaxy luminosities for the passive evolution 
of the stellar population, and cosmological dimming, we find that the
galaxy mass function has evolved little over this time.  
A factor of 2 evolution is only marginally consistent with the data
($\sim$2.5$\sigma$).

However, this level of uncertainty means that our data cannot yet
completely rule out the heirarchical model.
If we could fix the mass function normalisation using the total
(ie., dark plus stellar mass) of the clusters the evolution in galaxy
mass predicted by the theoretical models would be much clearer.
Unfortunately the clusters we
have studied have no reliable mass estimates that we can use to
independently secure the LF normalisation (the only measure available is a
velocity dispersion of 397 km s$^{-1}$ for the 3C336 cluster, measured from
eight cluster members: Steidel et al.\ 1997). This is an important area 
for future work.

There is no significant difference in the stellar mass function between
the 3c336 and Q1335 clusters ($<1.5\sigma$).
The cluster to cluster scatter in the theoretical models is also small
($\Delta$ $R$(giant/dwarf)=0.05). In an individual cluster, the measured
luminosity function is dependent on the background subtraction.
However, even if we artificially change the background level by a factor 2,
the resulting change in the giant/dwarf ratio is found to be less than
0.1. It therefore seems that our conclusion is not significantly
affected by the field-to-field variation of the background.

Finally, there are uncertainties in the theoretical model predictions. For this
reason it has been useful to check that we obtain similar results for 
both the Durham and Mitaka models. In addition,
the simulations allow us to check that cluster to cluster variation is not 
significant.  A concern when comparing theoretical and observational
results is the radius over which the comparison is made. The wide field
of our observations ($r_c$$\lsim$1~Mpc) means that they include a large
fraction of the cluster virial radius, and are thus directly comparable
to the observational results. It is interesting to note that the 
discrepancy between our results and the deficit of faint ($<M_{*}$+1.5)
galaxies by Kajisawa et al. (2000), might result from this type of mass
segregation.

\section{Conclusions}

We have presented the near-infrared LF of galaxies in the two high confidence 
$z\sim1$ cluster candidates based on deep wide-field imaging.
A third cluster was observed, but has then been excluded because
of contamination by a foreground system.
We find that the $K_s$-band LF at $z\sim1$ is 
consistent with pure luminosity evolution with constant stellar
mass according to the passive evolution with old age ($z_f\sim2$),
if compared to the lower redshift counterparts, such as 2MASS/LCRS LF for
local clusters.
We have transformed our LF at $z\sim1$ to a stellar mass function by
correcting for the star formation contribution to the $K_s$-band light
as estimated from the $J-K_s$ colours.
The derived stellar mass function is found to be very similar to the
one for the 2MASS/LCRS local clusters. In particular, massive galaxies 
($>$$10^{11}$M$_{\odot}$) make up a significant fraction of the galaxy
population. This result presents a challenge 
for current hierarchical galaxy formation models which tend to
under-predict the fraction of massive galaxies compared to our observed data.

The next steps are threefold.
One is to increase the sample of clusters at similar redshift ($z\sim1$)
and to have better statistics on the derived stellar mass function at this
epoch. The second is to extend this analysis
to higher redshift ($z\gg1$) to identify the epoch at which the
massive galaxies are assembled.  
The third is to extend this analysis along an environmental axis.
Galaxy-galaxy mergers are likely to be most effective in the group
environment rather than the cluster cores where velocity dispersion is
too high (Binney \& Tremaine 1987).
Therefore, it may be the case that the mass growth can be clearly 
seen at lower redshifts in these lower density regions.

\section*{Acknowledgements}

We are grateful to Dr. Masa Nagashima, Carlton Baugh, and Sarah Reed
for providing us with the model predictions for stellar mass functions
of galaxies from their semi-analytic models of galaxy formation,
and also for useful discussion about the comparison with the observed data.
We acknowledge to the referee, Dr. Arag\'on-Salamanca, for his useful
comments which have improved the clarity of the paper.
We thank Ian Smail for providing us with scripts for efficient observation
with INGRID and data reduction in IRAF.
TK thanks the Japan Society
for the Promotion of Science for support through its Research
Fellowships for Young Scientists between 1999-2001 during which
these INGRID data were taken. RGB thanks PPARC and the Leverhulme Trust for 
their support. We also acknowledge the Daiwa-Adrian Prize 2001
given by The Daiwa Anglo-Japanese Foundation.

\end{document}